\def\year{2022}\relax
\documentclass[letterpaper]{article} 
\usepackage{aaai22}  
\usepackage{times}  
\usepackage{helvet}  
\usepackage{courier}  
\usepackage[hyphens]{url}  
\usepackage{graphicx} 
\usepackage{natbib}  
\usepackage{caption} 
\usepackage{booktabs}
\usepackage[export]{adjustbox}
\usepackage{graphbox}
\usepackage{subcaption}
\usepackage{enumitem}

\DeclareCaptionStyle{ruled}{labelfont=normalfont,labelsep=colon,strut=off} 
\usepackage{algorithm}
\usepackage{algorithmic}
\usepackage{amsmath}
\usepackage{amsfonts}
\usepackage{subcaption}
\usepackage{booktabs}
\usepackage{xcolor}

\usepackage{newfloat}
\usepackage{listings}
\floatstyle{ruled}
\newfloat{listing}{tb}{lst}{}
\floatname{listing}{Listing}
\title{How Effective Are Large Language Models in Community Diagnosis? Examining Eating Disorders in Online Spaces}
\title{Large Language Models Help Reveal Body Image Concerns in Online Eating Disorders Communities}
\title{Large Language Models Help Reveal Unhealthy Diet and Body Concerns in Online Eating Disorders Communities}

\author {
    Minh Duc Chu,
    Zihao He,
    Rebecca Dorn,
Kristina Lerman
}

\affiliations {
    USC Information Sciences Institute\\
    \{mhchu, zihaoh, rdorn\}@usc.edu, lerman@isi.edu
}

\begin{document}

\maketitle

\begin{abstract}
Eating disorders (ED), a severe mental health condition with high rates of mortality and morbidity, affect millions of people globally, especially adolescents. The proliferation of online communities that promote and normalize ED has been linked to this public health crisis. However, identifying harmful communities is challenging due to the use of coded language and other obfuscations.  To address this challenge, we propose a novel framework to surface implicit attitudes of online communities by adapting large language models (LLMs) to the language of the community. We describe an alignment method and evaluate results along multiple dimensions of semantics and affect. We then use the community-aligned LLM to respond to psychometric questionnaires designed to identify ED in individuals. We demonstrate that LLMs can effectively adopt community-specific perspectives and reveal significant variations in eating disorder risks in different online communities. These findings highlight the utility of LLMs to reveal implicit attitudes and collective mindsets of communities, offering new tools for mitigating harmful content on social media.


\end{abstract}

\maketitle

\section{Introduction}

\textcolor{red}{[\textbf{Warning: This paper discusses eating disorders, which some  may find distressing.}]}


\noindent Online communities have emerged as important spaces for socializing, sharing information, validation, and emotional support~\cite{ellison2007benefits,he2024reading}. However, maintaining safe, welcoming, and inclusive online communities is challenging for several reasons. Malicious actors and others behaving in bad faith can derail discussions or engage in harmful speech. Although platforms have made strides in moderating online interactions, notably in identifying speech that violates community norms, such as hate speech and personal attacks~\cite{jhaver2019human,rajadesingan2020quick}, obstacles remain in creating safe online spaces. One type of online users that are hard to moderate are those that radicalize other individuals by indoctrinating them into extreme ideologies~\cite{schmitz2022quantifying}, conspiracy movements~\cite{wang2022identifying}, or glorify harmful behaviors, including self-harm~\cite{Goldenberg2022ncri}. Such community members evade moderation by using coded language and misspellings to obfuscate their goals~\cite{chancellor2016thyghgapp,cobb2017not}. We address this challenge by proposing a novel framework that combines the text generation capabilities of large language models with validated psychometric instruments to probe the collective mindsets of online communities to identify unhealthy beliefs.

We demonstrate the utility of our approach by analyzing online eating disorders (ED) communities. 
ED are serious mental health conditions characterized by obsessive thoughts and unhealthy behaviors around food, eating, and body size. The condition, which includes anorexia, bulimia, binge eating disorder, bigorexia, and other related mental health conditions, affects 24 million people in the US ~\cite{vanHoeken2020review} and leaves sufferers 
severely compromised due to medical and functional complications.  
Online communities have been linked to ED. Since the Web's early days, online communities that glorify thinness and promote anorexia as an aesthetic rather than a serious medical condition have proliferated~\cite{Ging2018}, and social media has only increased their reach and scope~\cite{chancellor2016quantifying,lerman2023radicalized}. While such communities provide validation, a safe space to vent, and emotional support to individuals struggling with ED~\cite{oksanen2016proanorexia, YeshuaKatz2013stigma}, they may ultimately harm them by normalizing disordered behaviors and ideas and delaying recovery \cite{pater2016hunger, pater2017defining, chancellor2016recovery}. 

Given the dual nature of these communities--offering both support and potential harm--it is crucial to develop methods that can accurately assess and monitor their impact. Screening online users individually for ED through traditional surveys is costly, time-consuming, impractical, and potentially unethical. However, the vast amount of data generated by online communities offers a promising alternative. Social media posts from online communities may \textit{collectively} reveal their mindsets, providing insights into the motivations and attitudes of their members. 

In this paper, we collect 2.6M tweets discussing ED and related topics spanning a period from 2022 to 2023. We construct a retweet network and identify organically-formed communities~\cite{blondel2008fast} that frequently interact with one another but rarely with outsiders. These communities discuss topics including 
\emph{Pro Eating Disorders}, \emph{Anti Eating Disorders}, \emph{Keto Diet}, \emph{Body Image}, \emph{Healthy Lifestyle \& Weight Loss}, and \emph{Weight Loss Drugs}. 




To probe communities' attitudes toward unhealthy beliefs about bodies and dieting, we use psychometric tools developed to screen individuals for ED~\cite{ellen2019}. These tools examine an individual's attitudes toward body shape and size, dieting history, and fears about weight through a series of targeted questions. To apply these psychometric tools to communities, we first create proxy representation for the collective mindset of the communities via large language models (LLMs).
Specifically, we align an open-sourced LLM called Llama-3 \cite{llama3modelcard} to each community using instruction tuning on the community's posts, thereby adapting it to the language---and the mindset---of the community. We evaluate the alignment along multiple dimensions, including community classification, toxicity, emotion analysis, and corpora embedding comparison, showing that the finetuned LLM authentically captures the community's ``voice''. 

Next, we administer an ED questionnaire to LLMs aligned with different communities. Analysis of LLM responses reveals that the two communities discussing ED are very different. One, which we call the Pro-Eating Disorder community has the highest risk for ED, meeting criteria for an unhealthy relationship with food, eating, and body image more than any other community. In contrast, the other community, the Anti-Eating Disorder community, which is critical of the diet culture, shows the lowest risk compared to other communities. 
These results highlight the significant differences in ED risk across different online communities and underscore the importance of targeted interventions. By identifying communities with higher risk levels, public health initiatives can be more precisely directed to those in need the most.


In summary, through our innovative approach of leveraging finetuned LLMs as proxies for online communities, we provide a scalable and cost-effective method for understanding the complex dynamics and potential risks associated with ED discourses on social media. Our findings not only shed light on the distinct attitudes and tendencies within different online communities but also pave the way for targeted interventions and public health efforts to address this growing mental health crisis.

\section{Related Work}
ED have complex biopsychosocial etiology, with contributing biological factors, such as genetics and infections~\cite{aman2022prevalence}, as well as psychological comorbidities, such as anxiety and perfectionism.
Social media fuels body image concerns through exposure to heavily edited images that promote the `thin ideal' or the `muscular ideal', thereby amplifying a key risk factor for developing depression and ED ~\cite{choukas2022perfectstorm, Tiggemann2018, Talbot2017, Griffiths2016}. 
Studies have shown that people compare themselves to idealized body images and as a result, feel worse about their appearance~\cite{Saiphoo2019, fardouly2016social,choukas2022perfectstorm}.
Social media can further fuel body image concerns through negative feedback to users' content and the images posted by others. In extreme cases, the negative feedback can manifest as body shaming and cyberbullying. 

\paragraph{Online Pro-ED Communities}
Pro Eating Disorders (Pro-ED) communities are online spaces, like blogs and online forums,  that promote ED as a lifestyle, not an illness. Pro-ED communities provide a venue for individuals with ED to share tips on losing weight and concealing weight loss from others, as well as ``thinspiration''/``fitspiration'' images of very thin bodies to motivate weight loss~\cite{Ging2018}.
Researchers have argued that pro-ED communities have both positive and negative effects on individuals with ED.
On the positive side, content analysis and qualitative studies revealed that these communities provide social support~\cite{juarascio2010pro} and a sense of belonging to individuals who often feel stigmatized and misunderstood~\cite{oksanen2016proanorexia,YeshuaKatz2013stigma}. Members can find empathy, encouragement, a safe space to vent, and information to help them better understand and manage their illness~\cite{McCormack2010}. 
On the negative side, pro-ED communities often promote unhealthy behaviors that can exacerbate ED, such as extreme calorie restriction and over-exercising. Members may compete with each other in weight loss, ask the group to hold them accountable to their weight loss goals, or find ``buddies'' to go through the same difficult periods of food restriction. Combined with celebrating the ``thin/muscular ideal'', these communities can increase psychological distress around body image and encourage disordered behaviors to persist~\cite{Mento2021}. 

Computer scientists have studied pro-ED communities by analyzing their language to identify harmful content or at-risk users. 
\citet{pater2016hunger} characterized hashtags and media content associated with ED-related posts on Tumblr, Instagram, and Twitter.  
\citet{chancellor2016post} built a lexical classifier to predict which posts will be taken down for violating Instagram's rules against self-harm. The authors combined statistical text analysis with clinician annotations to predict the severity of an individual's illness~\cite{chancellor2016quantifying}. The same authors also compared pro-recovery and pro-ED communities to show how people move from illness to recovery~\cite{chancellor2016recovery}. In contrast, 
rather than focus on individuals, we aim to identify the collective social dynamics mechanisms, which can be targeted to disrupt the growth of pro-ED communities. \citet{wang2018social} examined the retweet network of people discussing ED and their emotions, relying on the individual usage of hashtags to classify attitudes toward ED or recovery. Similarly, \citet{Wang2019} applied clustering algorithms to identify topics based on hashtags in Twitter conversations among users with ED and represent these interactions as a multilayer network to analyze the structural properties and temporal changes in communication patterns. In contrast, we leverage advanced language models to identify the community's attitudes toward mental health issues and body image concerns to gain deep, contextual insights into the aggregate linguistic patterns of the entire corpus of all users' full posts in each community. 

\paragraph{LLMs and Psychometric Tests}


\citet{binz} proposed adopting insights from cognitive psychology to understand LLMs more comprehensively. Their approach involves treating GPT-3 as a participant in psychological experiments to uncover the system's decision-making processes, reasoning abilities, cognitive biases, and other significant psychological traits. They demonstrated that evaluating machine behaviors by harnessing human psychometric tests in a principled and quantifiable manner can mitigate the limitations imposed by using human-curated benchmarks. \citet{coda2023inducing} showed that GPT-3.5 consistently generated robust responses to a widely used anxiety questionnaire, yielding anxiety scores higher than those of human participants. Through a systematic assessment of LLMs using a personality questionnaire, \citet{jiang2022evaluating} found initial evidence demonstrating the presence of personality within these models. While base models appear to respond to personality assessments inconsistently \cite{romero2023gpt, pellert2023ai}, adding persona instructions via in-context learning alongside questions from the personality questionnaire allowed PaLM \cite{chowdhery2022palm} to reliably emulate personality traits, similar to humans~\cite{serapio2023personality}.


\paragraph{Stanford-Washington University Eating Disorder (SWED) 3.0 Screener}


The Stanford-Washington University Eating Disorder Screener (SWED, \citet{Graham2019}) is a concise screening tool for identifying eating disorder behaviors and potential DSM-5 diagnoses. SWED incorporates questions from the Weight Concerns Scale (WCS,~\citet{wcs}), items from the Eating Disorder Examination-Questionnaire~\cite{Fairburn1994}, and the Eating Disorder Diagnostic Scale~\cite{Stice2000}. It has been widely used in research with both men and women \cite{ellen2019} and has even been incorporated into an accessible online screener tool \cite{neda-tool} by the National Eating Disorders Association \cite{neda} as part of its efforts to raise awareness about the condition.
Respondents receive feedback from the screener based on their answers, with pointers to treatment and resources.  

The Weight Concerns Scale (WCS), developed by \citet{killen1993}, is a brief, validated psychometric instrument designed to measure concerns about weight and body shape, fear of gaining weight, dieting history, and feelings of fatness. The scale was created to assess risk factors for ED and has demonstrated excellent stability and sensitivity to treatment differences. Longitudinal studies have shown that high WCS scores predict the onset of ED, making it a valuable instrument for identifying at-risk individuals \cite{killen1994, killen1996, jacobi2006}. Our work deploys this questionnaire to gain insights into body-related concerns of members of online communities.

\paragraph{LLM Alignment to Subgroups}
There is growing research on aligning language models to different subgroups to mimic their mindset and language.
Sub-population representative behavior models (SRBM) \cite{simmons2023large} emulate some characteristics of a particular subpopulation. 
Using SELF-INSTRUCT \cite{wang2023self} with seed instructions from the American Trends Panel, \citet{chen2024susceptible} show that LLMs can be ideologically manipulated to reflect the political opinions of specific subgroups.
To learn about political communities, \citet{jiang2022communitylm} propose communityLM by finetuning two GPT-2 \cite{radford2019language} models using causal language modeling on tweets from liberals and conservatives, and assessing the worldviews of the two groups using prompt-based probing of their corresponding finetuned models. \citet{he2024reading} extend communityLM by studying the mindset of organically-formed online communities and using message-passing to make use of the interactions between different communities. 

Instead of using base language models (such as GPT-2 \cite{radford2019language}), to leverage the strong language capabilities of large language models, we propose to finetune them using instructions following \cite{ouyang2022training}. 

\section{Identifying ED Communities in Online Discussions}

Twitter does not appear to moderate content related to ED, and as a result, pro-ED content has proliferated~\cite{Goldenberg2022ncri}. 
We gathered a large set of tweets about ED and related topics like weight loss, dieting, nutrition, etc. Our goal is to analyze these conversations to identify communities with unhealthy beliefs about bodies and food that are typically associated with ED. 

\subsection{Data Collection}
\label{sec:data}
We collected 2.6M tweets from 557K users on a Twitter query for ED-related keywords from October 2022 to March 2023. 
For keywords, we start with an existing set of terms that promote ED ~\cite{chancellor2016post,pater2016hunger}. This includes \textit{thinspo} (short-form for thinspiration), \textit{proana} (pro-anorexia), and \textit{promia} (pro-bulimia), among others. We altered the pre-existing list to remove spam terms yielding unrelated content, such as \textit{skinny}.  We expanded our query set to broader topics that are closely relevant to ED such as diet and weight loss including terms such as (\textit{ketodiet}, \textit{weightloss}, $\ldots$), and anti-diet culture (\textit{bodypositivity}, \textit{dietculture}, $\ldots$). A full list of keywords is featured in the Appendix. Information collected on posts includes user profile descriptions, timestamps, message content, and hashtags. 

\subsection{Community Detection}
\label{sec:res_echo_chamber}
To identify online communities, we construct a retweet network where each node represents a unique user and each undirected edge signifies one user retweeting another user at least once.
We use Louivain modularity maximization~\cite{Blondel_2008} to identify tightly-knit clusters of nodes, or communities, of users who frequently retweet each other.

Our community detection process identifies 402 communities. We focus on the 20 largest communities, which contain 71\% of the tweets and 40\% of the users in our dataset.  The statistics of these communities are shown in the Appendix, where community sizes drop quickly from 62K users to 3K users.

\begin{figure}[ht]
    \centering
    \includegraphics[valign=c,width=\linewidth]{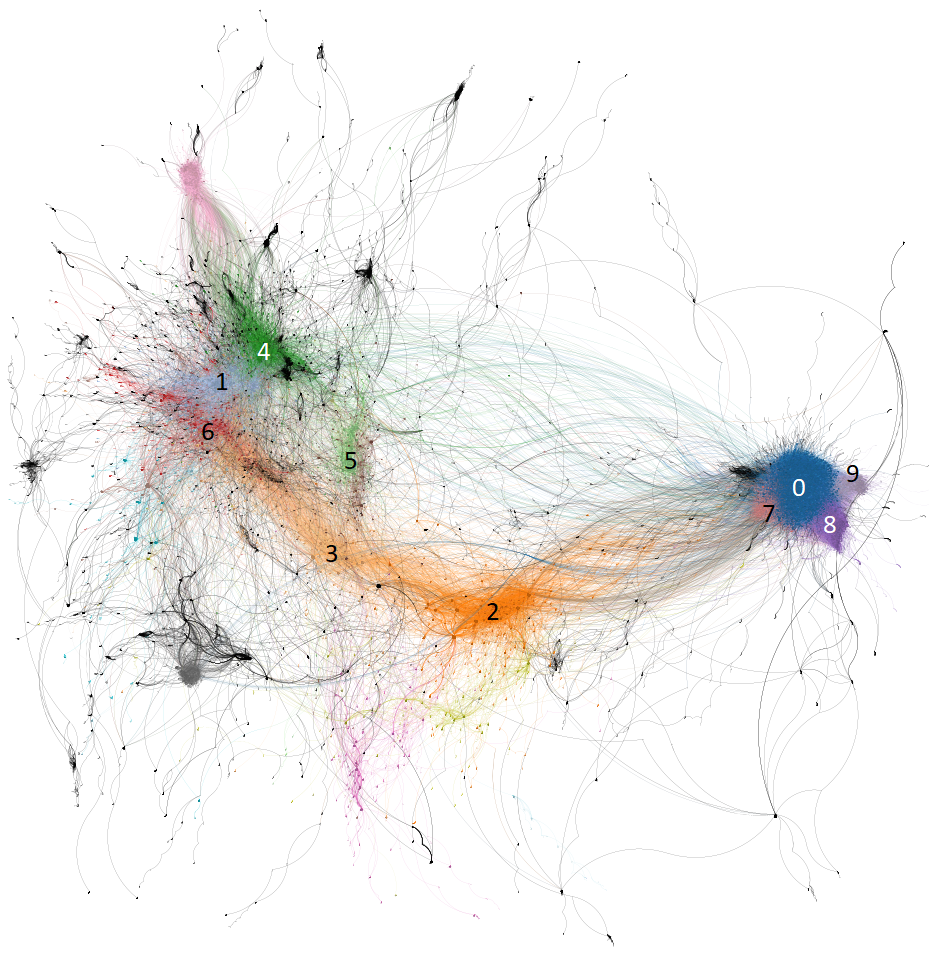}

\caption{Communities in the retweet network. User network showing retweets between individual users. Colors correspond to different communities identified by the Louvain method.}
\label{fig:rt_network}
\end{figure}


Figure \ref{fig:rt_network} shows the retweet network of users in our data, and we color the users from the top 10 largest communities.
Community names are integer-coded based on size, with larger numbers indicating smaller user populations. 
Most of the retweets are self-loops, suggesting that communities are \textit{echo chambers} of like-minded users. These echo chambers are the substrate via which harmful narratives may be amplified and spread. 
Communities 7, 8, and 9 are placed near community 0 in the network visualization due to the many links between them.

To profile the issues discussed within each community, we provide a random sample of 200 posts from each community to GPT-4 with the prompt: ``\emph{Given this list of posts, summarize the main ideas in 1 sentence}''. Starting with different random samples of posts leads to substantially the same generated summaries\footnote{We admit that such a way to profile communities is rough. However, by aligning LLMs which we will show in subsequent sections, we delve deeper into the mindset of the communities.}.

\begin{table*}[ht]
\centering
\footnotesize
\begin{tabular}{p{0.15\linewidth}|p{0.66\linewidth}|l} \hline
\textbf{Grouped Comm. Name}                           & \multicolumn{1}{c|}{\textbf{Community Description}}                                    & \textbf{Comm. ID}                           \\ \hline
Pro Eating Disorder                                                         & This community revolves around the online eating disorder community (edtwt), sharing tips, thinspo (thin inspiration), meanspo (mean inspiration), fasting strategies, and discussing body image and weight loss goals, often in a way that promotes disordered eating behaviors.                                              & 0,7,8,9      \\
\textcolor{gray}{Keto \& Diet}                                                                & \textcolor{gray}{This community focuses on a range of topics related to ketogenic diets, weight loss, metabolic health, and low-carb recipes, with discussions on the effectiveness of keto for various health conditions, debates on prescribing obesity drugs to children, and personal testimonials about the benefits of a keto.}            & \textcolor{gray}{1,15,16,18}   \\
Body Image                                                                  & This community dives into a variety of personal updates, including fitness activities, body positivity, nudism, modeling, and social interactions, with some tweets promoting content or expressing motivational thoughts.                                                                                                     & 10           \\
\textcolor{gray}{Anti Eating Disorder}                                                        & \textcolor{gray}{This community expresses strong negative sentiments towards "edtwt" (presumably "eating disorder Twitter"), criticizing it for being toxic, fatphobic, and harmful, with calls to abolish it and stop interacting with its content.}                                                                                            & \textcolor{gray}{2}            \\
Healthy Lifestyle \& Weight Loss & This community covers a variety of health and wellness topics, including weight loss methods, dietary plans, fitness advice, healthy eating, keto diet, fasting, moxibustion, and motivational messages for maintaining a healthy lifestyle.                                                                                   & 4,13,17      \\
\textcolor{gray}{Weight Loss Drugs}                                                           & \textcolor{gray}{This community discusses the controversial use of the diabetes drug Ozempic for weight loss, the impact of its shortage on diabetic patients, the cost of the medication, and related topics such as body positivity, keto diets, and the role of influencers and celebrities in promoting certain health trends and products.} & \textcolor{gray}{3,6,19}  \\ \hline    
\end{tabular}
\caption{Summary of posts in the communities with GPT-4. Similar communities are merged.}
\label{tab:comm_summaries}
\end{table*}

Summaries of the ten largest communities are shown in Table~\ref{tab:comm_summaries}, and their word clouds are shown in Figure \ref{fig:wordclouds} in the Appendix. Based on the summaries and the word clouds, we assign six unique labels to them: \emph{Pro Eating Disorder}, \emph{Keto and Diet}, \emph{Body Image}, \emph{Anti Eating Disorder}, \emph{Healthy Lifestyle and Weight Loss}, \emph{Weight Loss Drugs}, and \emph{spam} (not included). 
GPT-4 identifies clusters 0, 2, 7, 8, and 9 as belonging to the eating disorder community on Twitter. This is confirmed by word-level analysis, which shows that clusters 0, 2, 7, 8, and 9 contain a large share of posts mentioning ``edtwt'' a term used by members of the ED community on Twitter to self-identify and tag posts. Interestingly, clusters 0, 7, 8, and 9 are all devoted to pro-ED discussions, although cluster 8 contains many Spanish language posts and cluster 9 contains posts in Portuguese, suggesting these are international pro-ED communities. They are also placed tightly close to each other in Figure \ref{fig:rt_network}. Community 2, although also dedicated to eating disorder discussions, is well separated from the rest. This community takes a critical---anti-ED---stance on ED, as seen from the summary in Table~\ref{tab:comm_summaries}.


The remaining large communities form a loosely connected cluster where no community is isolated from the rest. These communities share posts about diet and weight loss but they are not as insular as the pro-ED community is. Community 1 discusses the risks and benefits of the keto diet; communities 3 and 6 focus on issues surrounding the use of weight loss drugs like Ozempic and Wegovy; Community 4 looks at issues of healthy lifestyle and weight loss, while Community 5 covers body image topics, like body positivity and self-acceptance. Communities 11, 12, and 14 are on other random issues not relevant to ED, as can be observed from word clouds (Figure \ref{fig:wordclouds} in Appendix), and thus we exclude them in our subsequent analysis.

Although we do not have all the retweets made by the users but only those containing our search query terms, the patterns of connectivity that we observe suggest that they faithfully capture the structure of attention in the discussions of these topics. 
There is a notable lack of retweet interactions between members of the eating disorder communities (0, 2, 7, 8, 9). This suggests that {eating disorder communities on Twitter form an echo chamber}, which traps users within toxic mindsets, delaying recovery. The only exception is Community 2 which appears to interact with the ED community but holds critical views of it. 

Drawing on insights from the GPT-4 summaries and word clouds, we note significant thematic and content overlaps among the Louvain-detected communities (later referred to as \textbf{retweet communities}) in the retweet network, as identified by integers in the ``Comm. ID'' column of Table \ref{tab:comm_summaries}. 
We therefore consolidate related communities into larger groups (\textbf{grouped communities}), as outlined in the ``Grouped Comm. Name'' column of Table \ref{tab:comm_summaries}. Intriguingly, in the retweet network (Figure\ref{fig:rt_network}), communities grouped by similar content are close together in the retweet network. For instance, the grouped communities on ED, including Communities 0 and 7 (pro-ED), along with Communities 8 and 9 (pro-ED in Portuguese and Spanish), are not only network-adjacent but also interconnected through a significant number of links, highlighting their thematic alignment. \textbf{In our following discussions, we will only focus on grouped communities} (excluding the grouped community \emph{spam}).

\section{Aligning LLMs to Communities}

\begin{figure*}[ht]
    \centering
    \includegraphics[width=0.7\linewidth]{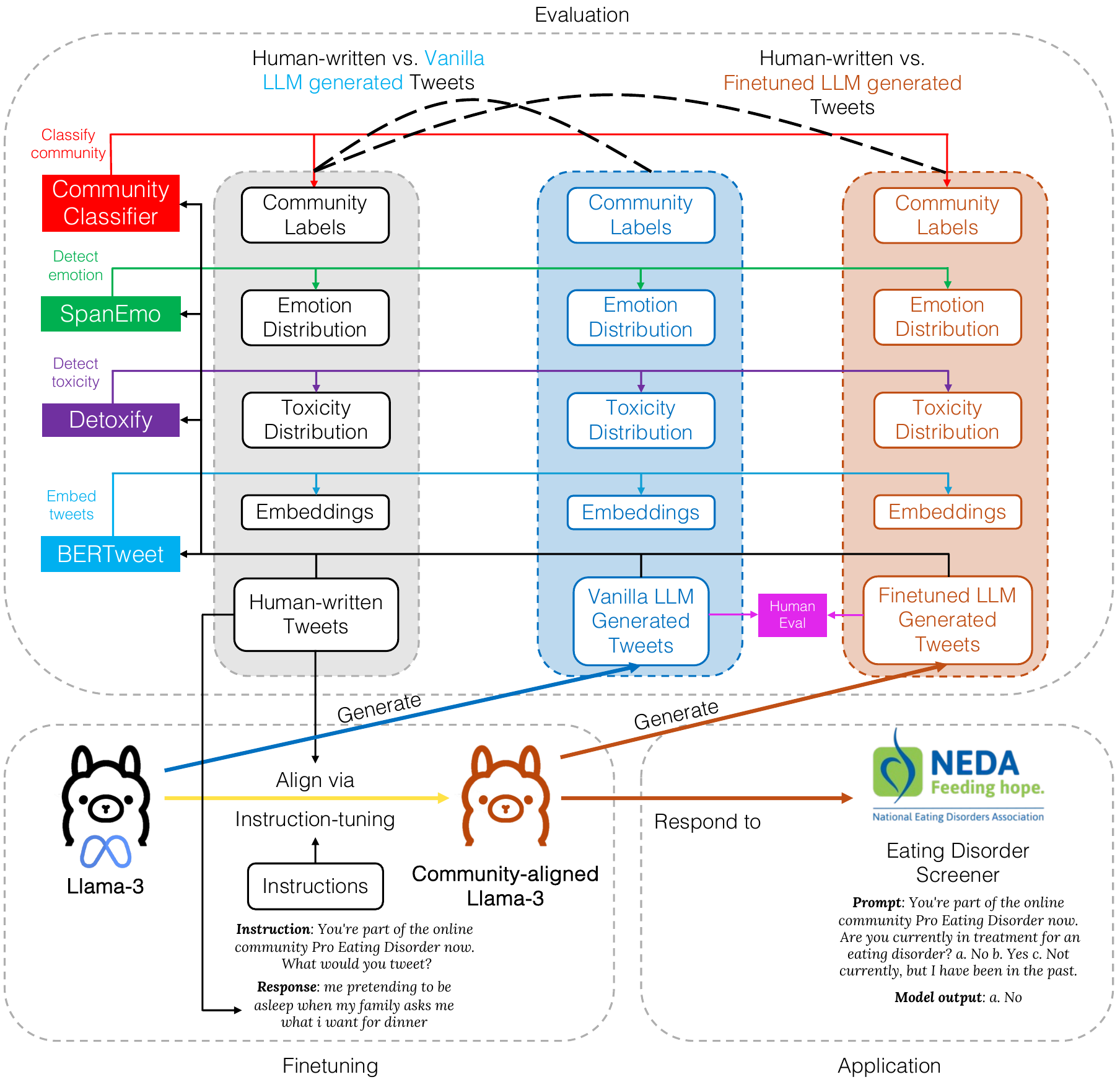}
\caption{The framework of our method. 
(1) We align an LLM (Llama-3) to the language and mindset of an ED community. The alignment is achieved by finetuning the LLM to generate tweets written by users in the community by following instructions. 
(2) To prove that the alignment is effective, we focus on three sets of tweets: $\alpha$. human-written tweets, $\beta$. vanilla (unfinetuned) LLM-generated tweets, and $\gamma$. finetuned LLM-generated tweets. We show that $\gamma$ is closer to $\alpha$ than $\beta$ is, from the following aspects: (a) A classifier trained to classify the community origins of $\alpha$ performs equally well on $\gamma$, but not on $\beta$; (b) the emotion and toxicity distributions of $\gamma$ are much closer to that of $\alpha$ compared to that of $\beta$ are; (c) the embeddings of $\gamma$ are closer to that of $\alpha$ in the embedding space than that of $\beta$ are; (d) the human annotator decides that $\gamma$ is more aligned to underlying distribution of $\alpha$ than $\beta$ is.
(3) As the LLM is aligned with the community and can speak as a typical individual in the community, we administer an eating disorder questionnaire to it and aim to screen the community for ED. 
}
\label{fig:framework}
\end{figure*}



We describe and evaluate a method to align language models with community-specific social media content through instruction tuning \cite{jiang2022communitylm, he2024reading, alpaca}. By finetuning the LLM on the posts within the community, we hope that the LLM learns to capture not only the language but also the community's collective mindset. 
 
\subsection{Constructing Instruction-Response Pairs}
\label{sec:inst-resp-pairs}
We use the tweets from each community to construct demonstrations (instruction-response pairs) for finetuning the LLM, where the instruction is on tweet generation, and the response is the tweet. The demonstrations enable us to align the LLM to the language (both semantics and style)---and the mindset---of each community.
We preprocess tweets by removing unnecessary information like URLs, mentions, hashtags, and emojis, and we filter out retweets and comments.
To ensure high-quality data, we compute the perplexities of the tweets using BERTweet \cite{nguyen2020bertweet} and select a maximum of 10K highest quality (i.e., lowest perplexity) tweets. If there are fewer than 10K tweets from a community, we keep all the tweets. The numbers of tweets from the community \emph{Pro Eating Disorders}, \emph{Keto \& Diet}, \emph{Body Image}, \emph{Anti Eating Disorders}, \emph{Healthy Lifestyle \& Weight Loss}, and \emph{Weight Loss Drugs} are 10K, 10K, 3.3K, 2.9K, 10K, and 10K respectively.
Each selected tweet is then paired with an instruction from our instruction pool (Table \ref{tab:inst} in the Appendix). 
For each community, in addition to the demonstrations on tweet generation, we augment them with the 52K Alpaca \cite{alpaca} demonstrations that cover a wide range of different tasks, to retain the instruction-following capabilities of the LLM and not restrict it to just tweet generation.



\subsection{Instruction Tuning LLMs}
We use Llama3 \cite{llama3modelcard}, which is an open-source LLM, and align it to the language of a community by instruction tuning. The finetuned model can act as a proxy representation for a collective set of users within the community, whose mindset we can probe to understand the complex dynamics of the community online discourses and identify ED ideologies. 
For each community, we finetune the LLM on 4 Tesla H100-80GB GPUs with batch size 8 for 3 epochs. The finetuning takes about 3 hrs.

\subsection{Measuring Alignment}
To demonstrate that the finetuned LLMs are effectively aligned with their respective communities, we first prompt the LLMs to generate tweets using the same instructions as in finetuning. However, to diversify the LLM generations, we compile a set of topics relevant to eating disorder discussions, such as \emph{thinspo}, \emph{fitspo}, and \emph{bonespo} (refer to the Appendix for the complete topics), and prompt LLMs to generate tweets on these topics. An example instruction is ``You're part of the online community \{community\_name\} now. What would you tweet \textbf{about fasting}?''. We empirically observe that without specifying the topics, the LLM would generate semantically uniform responses. For each topic, the LLM generates 400 responses. 
For the vanilla LLM, as it is not aligned to any community, simply specifying the community name is not sufficient for it to grasp its mindset. Therefore, when prompting the vanilla LLM to generate tweets, we further provide the community descript by GPT-4 (as shown in Table \ref{tab:comm_summaries}) in the prompt. Please refer to the Appendix for the prompt template.
Even though the original tweets may not fully cover these topics, after aligning with the community, the LLM can express opinions on these topics.
To evaluate alignment, we compare generated tweets from both vanilla and finetuned LLMs with the human-written tweets in four key dimensions: 1) community classification, 2) emotion and toxicity analysis, 3) tweet embedding comparison, and 4) human evaluation.

\begin{figure}[ht]
\centering
\includegraphics[width=0.46\textwidth]{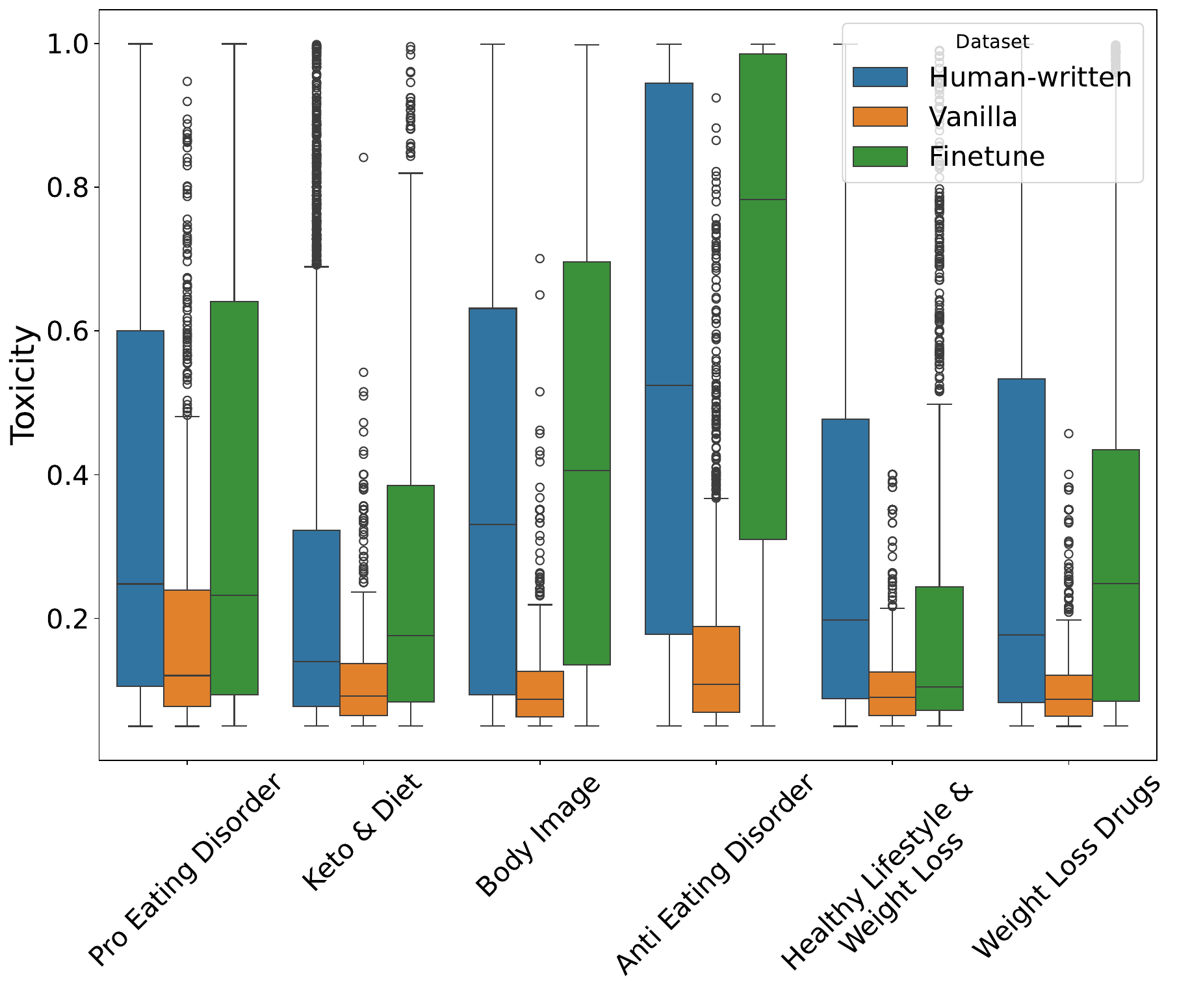}
\caption{Toxicity distribution across different communities of human-written posts, vanilla LLM generated posts, and finetuned LLM generated posts.
}
\label{fig:toxicity}
\end{figure}

\begin{figure}[tbh]
    \centering
    
    \begin{subfigure}[b]{\linewidth}
        \centering
        \includegraphics[width=0.75\textwidth]{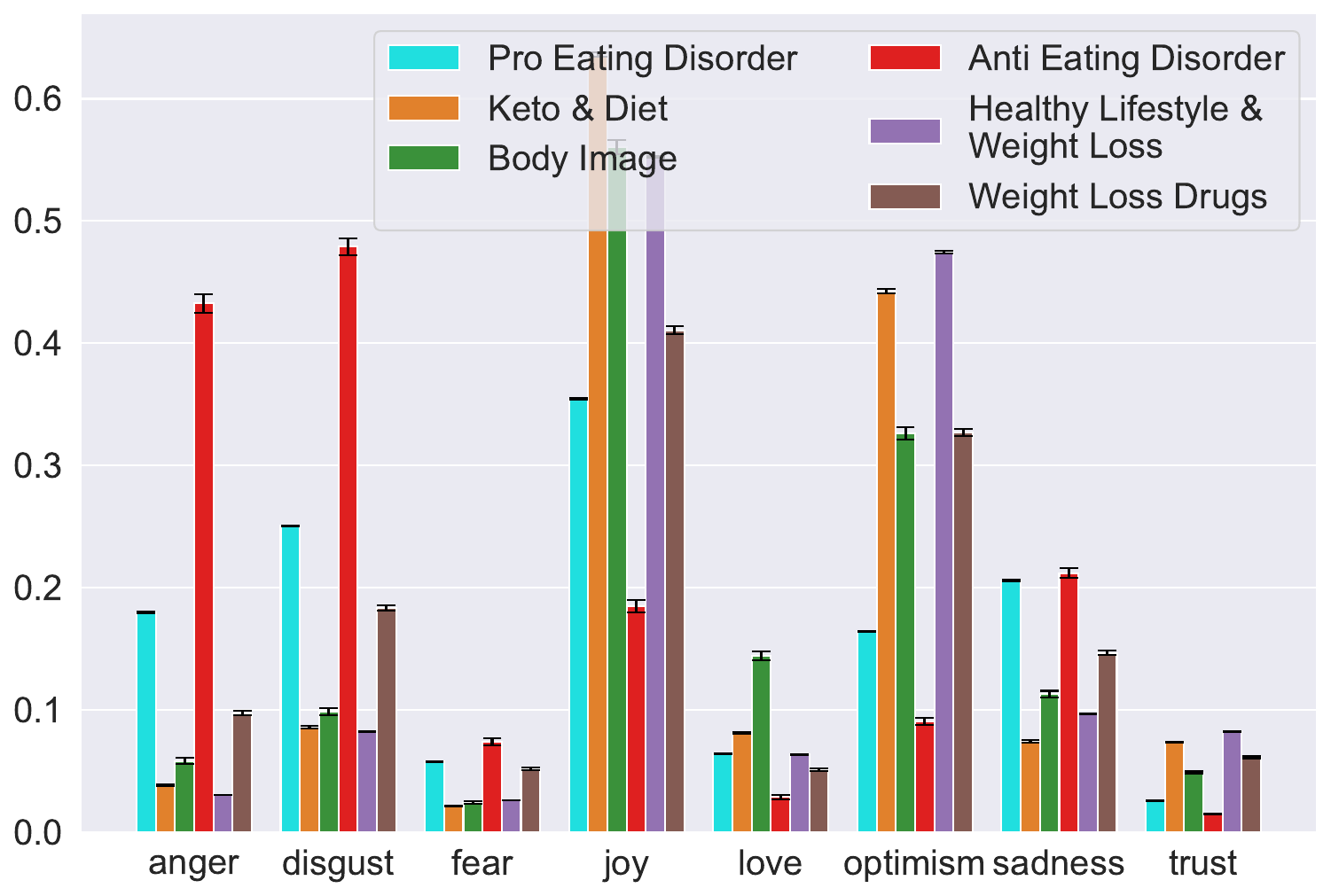}
        \caption{Human-written posts}
        \label{fig:emo-og}
    \end{subfigure}
    \hfill

    \begin{subfigure}[b]{\linewidth}
        \centering
        \includegraphics[width=0.75\textwidth]{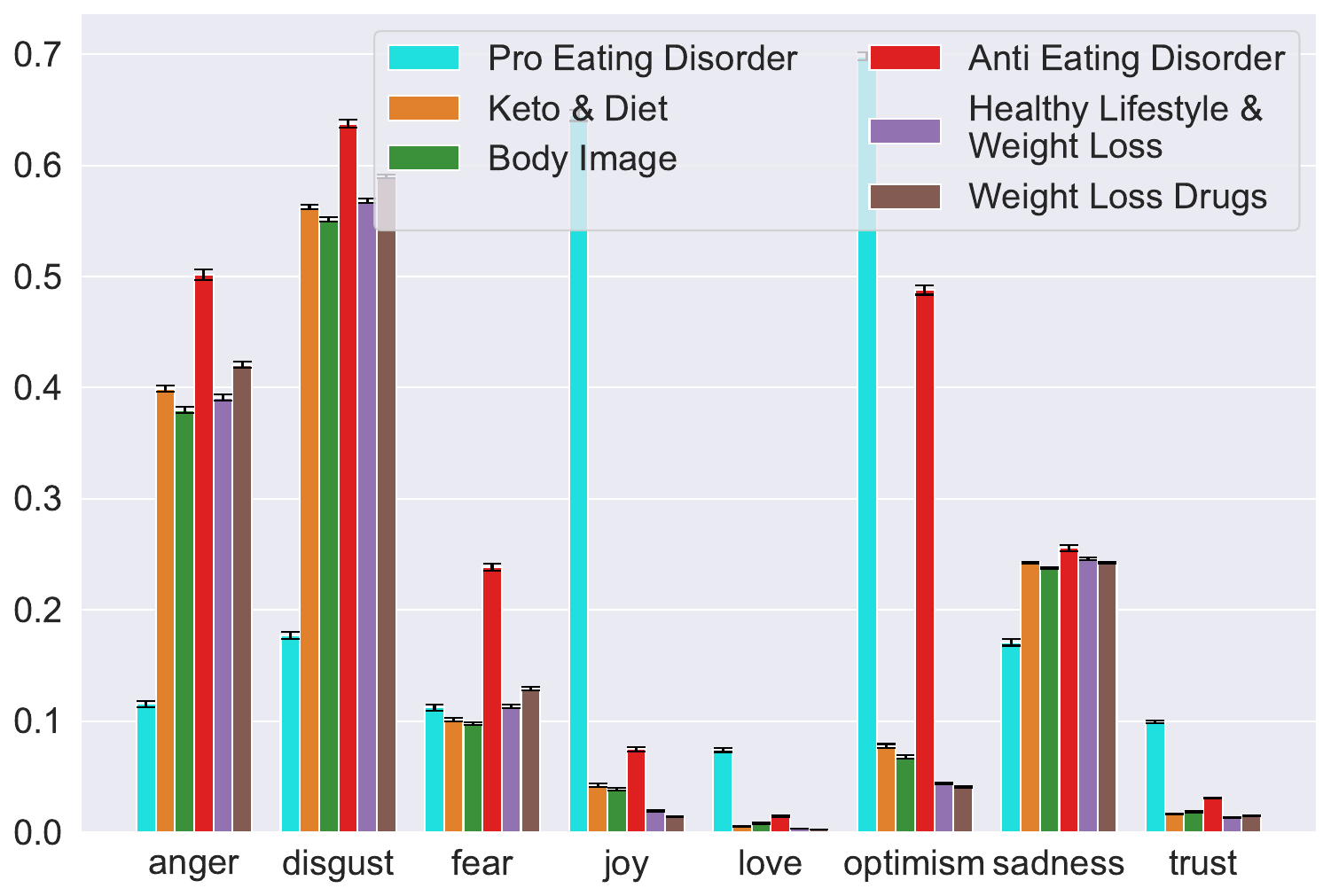}
        \caption{Vanilla LLM generated posts}
        \label{fig:emo-base}
    \end{subfigure}
    \hfill

    \begin{subfigure}[b]{\linewidth}
        \centering
        \includegraphics[width=0.75\textwidth]{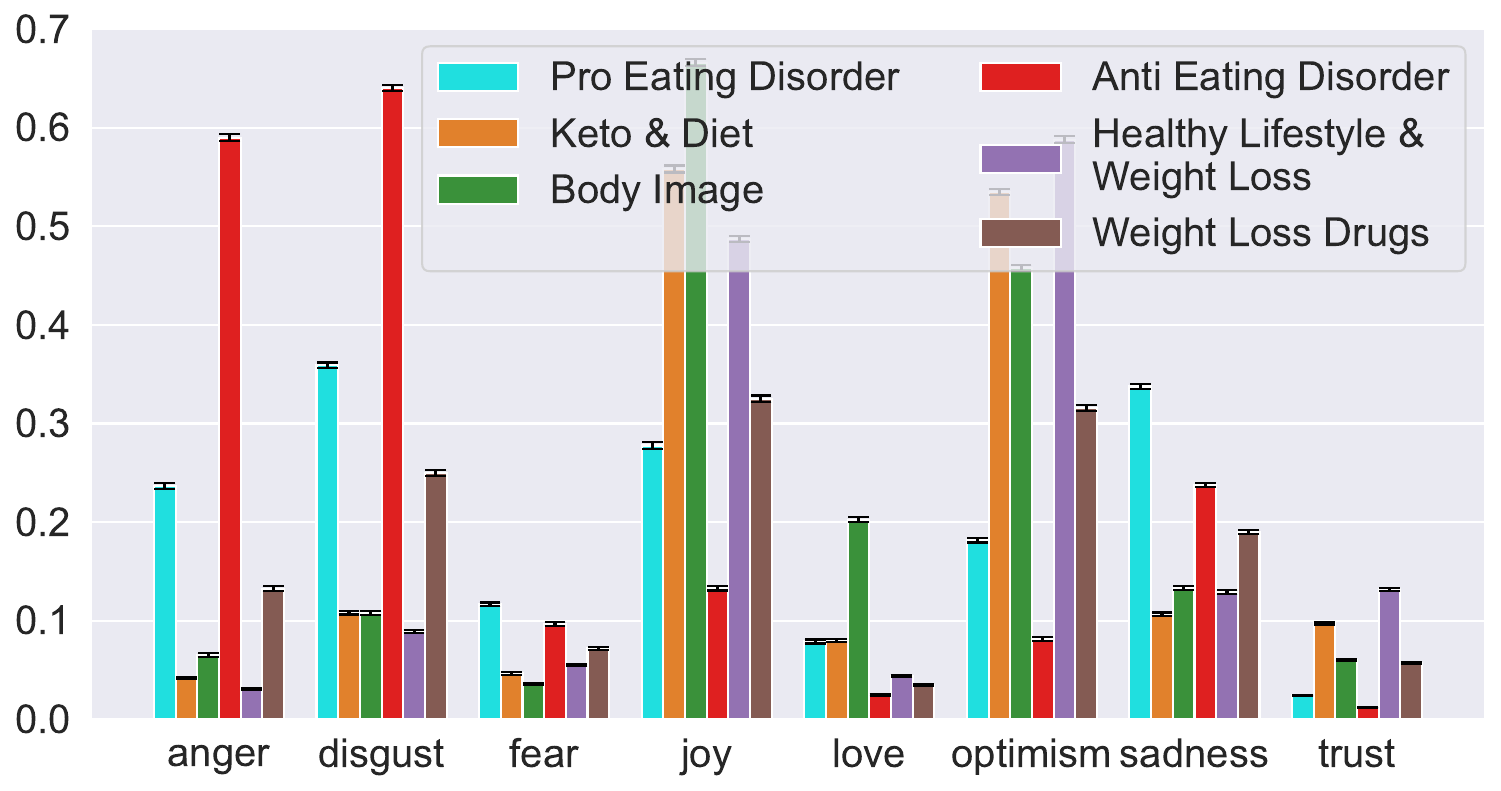}
        \caption{Finetuned LLM generated posts}
        \label{fig:emo-ft}
    \end{subfigure}
    
    \caption{Emotion distributions from different communities of (a) human-written posts, (b) vanilla LLM generated posts, and (c) finetuned LLM generated posts.
    }
    \label{fig:emo-dist}
\end{figure}

\paragraph{Community Classification}
We train a classifier to determine the community from which a tweet originated. We achieve this by finetuning Llama-3 using instructions with the following format ``\emph{From these communities: Pro Eating Disorder, Keto \& Diet, Body Image, Anti Eating Disorder, Healthy lifestyle \& Weight Loss, and Weight Loss Drugs; which community does this Tweet belong to? \{Tweet\}}'', with the expected response being the community name.
We sample 3,000 original tweets from each community and construct 18,000 demonstrations for finetuning. We train the classifier using 95\% of the demonstrations and use the remaining 5\% to test, leading to a test accuracy of 0.8.
Using this classifier, we classify the vanilla LLM-generated tweets and finetuned LLM-generated tweets, leading to an accuracy of 0.27 and 0.81, respectively. These results indicate that the classifier trained on original tweets accurately recognizes the tweets generated by the finetuned LLM. However, it performs poorly on the tweets generated by the vanilla LLM, demonstrating that the finetuned LLMs better capture community-specific linguistic characteristics.

\paragraph{Emotion and Toxicity Analysis}
We analyze tweets using detoxify~\cite{Detoxify}, a state-of-the-art library for detecting toxicity in text and a popular metric of harmful online communities~\cite{rajadesingan2020quick, sheth2022defining}. Although detoxify returns values between 0 and 1, for clarity and to avoid noise, we only include tweets with toxicity levels equal to or greater than 0.05. Figure \ref{fig:toxicity} shows the distribution of toxicity scores of human-written tweets, vanilla LLM-generated and finetuned LLM-generated tweets. 
The toxicity distribution of the finetuned LLM-generated tweets more closely matches that of the human-written tweets compared to the vanilla LLM-generated tweets, and tweets from the anti-ED community have the highest toxicity.

We also analyze the distributions of emotions of the real-world posts, posts generated by the vanilla LLM and those generated by the finetuned LLM. Emotions are classified using SpanEmo \cite{alhuzali2021spanemo}. The emotional tone can reveal whether the content is supportive, triggering, or neutral, which is essential for capturing the nuances of the community's interactions \cite{chmiel2011collective}. 
As shown in Figure \ref{fig:emo-dist}, the distribution of emotions expressed in tweets generated by the finetuned LLM is closer to the distribution of emotions in real-world tweets.


\paragraph{Corpora Embedding Comparison}
We compute the embeddings of the original tweets, vanilla LLM-generated tweets, and finetuned LLM-generated tweets using BERTweet \cite{nguyen2020bertweet}. We then measure the distance between these embeddings using the Fréchet Inception Distance (FID) \cite{heusel2017gans}. This metric provides a quantitative measure of how closely the generated tweets resemble the original tweets in terms of their semantic content. 
We implement it using the IBM comparing-corpora package \cite{kour2022measuring}. $FID(\text{human\_tweets}, \text{vanilla\_tweets})$ and $FID(\text{human\_tweets}, \text{finetuned\_tweets})$ for different communities are shown in Table \ref{tab:fid}. We see that $FID(\text{human\_tweets}, \text{vanilla\_tweets})$ is much smaller than $FID(\text{human\_tweets}, \text{finetuned\_tweets})$. 

\begin{table}[ht]
\addtolength{\tabcolsep}{-3.0pt}
\centering
\begin{tabular}{lcc}
\hline
\multicolumn{1}{c}{\textbf{Community}} & \textbf{\begin{tabular}[c]{@{}c@{}}$FID$(HT, VT)\end{tabular}} & \textbf{\begin{tabular}[c]{@{}c@{}}$FID$(HT, FT)\end{tabular}} \\ \hline
Pro Eating Disorder & 0.94 & 0.48 \\
Body Image & 1.91 & 0.37 \\
Keto \& Diet & 1.78 & 0.51 \\
Anti Eating Disorder & 1.14 & 0.52 \\
\begin{tabular}[c]{@{}l@{}}Healthy Lifestyle \&\\ Weight Loss\end{tabular} & 1.93 & 0.54 \\
Weight Loss Drugs & 1.96 & 0.40 \\ \hline
\end{tabular}
\addtolength{\tabcolsep}{3.0pt}
\caption{Fréchet Inception Distances (FID) between human-written tweets (HT) and vanilla LLM generated tweets (VT), and between human-written tweets and finetuned LLM generated tweets (FT). A smaller distance indicates more similarity.}
\label{tab:fid}
\end{table}

\paragraph{Human Evaluation}
In addition to automatic evaluation of the finetuned LLM in terms of the alignment with particular communities, we perform human evaluation. Specifically, an annotator with expertise with ED on social media is presented with 300 triplets, where a triplet consists of a community name, a vanilla LLM-generated tweet, and a finetuned LLM-generated tweet. Both tweets in a triplet are on the same topic. For each triplet, the annotator is asked to decide which tweet is more aligned with the underlying distribution of the community, by referring to the following characters:
mis/use of ingroup language, references to themes in underlying distribution (e.g. the Body Image community often references nudity), use of capitalization, and coherence of message.
In 274 out of 300 triplets, the annotator choose the finetuned LLM-generated tweet as a better match, indicating the finetuned LLM is more aligned with the community's characteristics.

\section{Screening Online Communities for ED} 




After aligning language models, we use them to probe online communities using existing established ED screening tools for evidence of body image concerns related to ED. 
Specifically, we prompt community-aligned LLMs to respond to questions in the SWED screening tool. SWED consists of 11 questions, multiple-choice and open-ended (see Appendix), 
covering demographics, height and weight, eating disorder behaviors, weight and shape concerns, and impairment. 
We focus on questions in SWED that are relevant in determining if the individual is at high risk of an eating disorder and evaluate model responses to them by using the following four criteria \cite{ellen2019}.


The first criterion is the WCS score - a metric that measures concerns about weight and shape. This criterion composes answers from questions 5 to 9 in SWED, maps the options to numerical values to a 0-100 scale, and calculates the average score across the 5 questions. The higher the WCS score is, the higher the risk factor for body and shape concerns.  

The next three criteria are coded as ``C1, C2, C3''. C1 is True if the model responds ``more/most important" (options c or d) to Q8 ``Compared to other things in your life, how important is your weight to you?". C2 is True if the model responds ``moderately afraid" or more (options c, d, or e) to Q6: ``How afraid are you of gaining 3 pounds?''. C3 is True if the model responds ``Yes'' at least 3 times to Q11: ``In the past 3 months, how many times have you done any of the following as a means to control your weight and shape: a. Made yourself throw up? b. Used diuretics or laxatives? c. Exercised excessively? d. Fasted?"

To account for randomness, for each question in the SWED questionnaire, the finetuned LLM generates 50 responses. The responses are parsed and aggregated using a majority vote for each question. Results, shown in Table \ref{tab:swed}, indicate that the \emph{Pro Eating Disorders} community exhibits the highest degree of concern about weight and shape, followed by the \emph{Keto \& Diet} community. Furthermore, both communities meet all three criteria signaling a high likelihood of eating disorder pathology, whereas responses of the \emph{Anti Eating Disorder} community are consistent with a low risk of ED. 

These findings align with our empirical observations of these communities. The content shared by the \emph{Pro Eating Disorders} community revolves around sharing tips to promote disordered behaviors and body dysmorphia. Conversely, the \emph{Anti Eating Disorder} community is critical of the diet culture and people who glorify ED. 
The relatively high risk score of the \emph{Keto \& Diet} community is a concerning indicator that this community may serve as a gateway to ED. 
In contrast, the Body Image community, which mostly posts about body positivity, has low scores, implying low risk for ED, as does the Healthy Lifestyle \& Weight Loss community. Although the latter focuses on Weight Loss, it appears to achieve this goal through healthy behaviors.



\begin{table}[ht]
\centering
\begin{tabular}{l|ccll}
\hline
\multicolumn{1}{c|}{\textbf{Community}}                                    & \textbf{WCS} & \textbf{C1} & \textbf{C2} & \textbf{C3} \\ \hline
Pro Eating Disorder                                                        & \textbf{45.0}   & T           & T           & T           \\
Keto \& Diet                                                               & 33.3         & T           & T           & T           \\
Weight Loss Drugs                                                          & 16.7         & F           & F           & T           \\ 
Body Image                                                                 & 15.0         & F           & F           & F           \\
\begin{tabular}[c]{@{}l@{}}Healthy Lifestyle \&\\ Weight Loss\end{tabular} & 13.3         & T           & F           & F           \\
Anti Eating Disorder                                                       & 13.3         & F           & F           & F           \\
\hline
\end{tabular}
\caption{Evaluation of the responses of different community-aligned LLMs using four criteria, including WCS, C1, C2, and C3.
``C'', ``T'', and ``F'' are short for ``criterion'', ``True'', and ``False'' respectively. For WCS, a higher score indicates a higher risk of an eating disorder. For C1, C2, and C3, the more positive responses, the higher the risk.}
\label{tab:swed}
\end{table}

\section{Conclusion}

Our study highlights the utility of LLMs in revealing the implicit perspectives of online communities related to ED. By aligning LLMs to specific community languages, we uncovered significant variations in the risk levels of different communities. Our findings reveal that pro-eating disorder communities exhibit attitudes associated with a high risk of ED condition, while anti-eating disorder and healthy lifestyle communities show much lower risks. These insights are consistent with the semantic analysis of tweets via GPT-4, demonstrating robustness. This approach provides a scalable and cost-effective method for public health officials to monitor and target interventions effectively. Overall, leveraging LLMs offers a powerful tool for addressing the complex issues of ED in digital spaces, enabling more precise and impactful public health strategies.

In the future, we plan to enhance our community alignment framework by integrating additional data modalities such as images, audio, and videos from diverse social media platforms, including TikTok and Reddit.  To better gain a comprehensive and subtle understanding of the complex and multifaceted narratives and accurately assess the mental health of ED communities, we need to include humans in the loop. For example, while some content may be well-intentioned, such as sharing personal experiences and coping strategies, it can inadvertently trigger and accidentally share ED tactics to vulnerable individuals. To address this, we aim to collaborate with psychiatrists, psychologists, and sociologists specializing in ED to develop a harm taxonomy tailored to the ED content on social media. Moreover, given the variability in content moderation policies across platforms and the creative ways users circumvent these policies \cite{bickham2024hidden}, such as using intentional misspellings, there is a critical need for a robust system. This system must adeptly handle these challenges to effectively support mental health and eating disorder interventions.

\section{Limitations}

\subsubsection{Complete Coverage of Eating Disorders.}
This paper looks at the discussions of ED in online communities. We focus on a conglomeration of ED, including bulimia nervosa, anorexia nervosa, and binge eating disorder. 
Besides ED, our dataset captures other discussions related to weight concerns, such as weight loss, diet, body positivity, etc.
The full list of keywords used for dataset compilation is given in the Appendix Table \ref{tab:keywords}.
Unfortunately, our data does not comprehensively represent all existing ED.
However, our methods ensure that if a large eating disorder community has some overlap with our keyword list, the community will be identified.


\subsubsection{Sensitivity to Prompt Perturbations.}
Small changes to prompts such as choice of output format can alter LLM performance on classification tasks \cite{salinas2024butterfly}. For this reason, we instruction-tune using a set of prompts as featured in Appendix Table \ref{tab:inst}.

\subsubsection{Dataset Bias.}
The anonymized version of our dataset may contain implicit biases reflecting societal prejudices or unequal representation of demographic sub-groups. 
More specifically, eating disorder symptoms have a history of being under-diagnosed in African American and Hispanic adolescent girls, in part due to stereotypical representation of ED being Caucasian adolescent girls \cite{gordon2002impact}.
This historical bias could be inadvertently learned by our model, resulting in discriminatory behavior. In our future work, we hope to evaluate the model’s fairness across different user groups, allowing us to properly mitigate dataset biases.

\subsubsection{Temporal Dynamics.}
Conversations on social media change drastically over time. We have collected insight into online behavior in a particular snapshot of time. This may not capture the full range of topics or events in the communities at hand.

\bibliography{references}

\clearpage
\section{Ethics Statement}

\subsubsection{Community-Level Diagnosis.}
Diagnosing psychiatric illness at the community level comes with the risk of falsely diagnosing some community members.  This could lead to unjust actions against users, such as unwarranted bans or removal of content. Additionally, approximating community behavior inherently excludes minority group members. Simultaneously, anorexia is one of the deadliest mental health disorders\footnote{https://www.state.sc.us/dmh/anorexia/statistics.htm} and participation in online pro-ED spaces heightens one's disease risk \cite{Mento2021}. By evaluating psychiatric illness on the community level, we can identify toxic communities, helping content moderation experts deploy proper interventions to promote healthy and safe online environments. We encourage the use of human moderators to review and validate the decisions made by our model, particularly in cases with low confidence scores.

\subsubsection{Topic Sensitivity and Privacy}
The sensitive nature of our topic means that our outputs could be misused, such as targeted advertising.
Additionally, our dataset includes some tweets that disclose deeply personal information such as medical diagnoses, weight information, and personal struggles. Many of these tweets are posted under the assumption of anonymous identity. By collecting these tweets, user-specific information may be pieced together thus de-anonymizing some users. For these reasons, we take precautions to anonymize the social media posts before feeding them to the language models. Additionally, researchers can be granted access to generated tweets upon detailed inquiry.


\subsubsection{Hallucination Risk.}
It is possible for our fine-tuned models to exhibit hallucination, generating incorrect or nonsensical information. Hallucination in the context of community alignment can lead to community misrepresentation. In future work, we hope to utilize some factual-based evaluation datasets to measure model hallucination.

\clearpage
\appendix

\section{Data Collection}

\subsection{Keyword Search Terms}
The keywords used for tweet collection are shown in Table \ref{tab:keywords}.

\begin{table}[h!]
\begin{tabular}{|l|}
\hline
\multicolumn{1}{|c|}{\textbf{Keywords}}           \\ \hline
\textit{\begin{tabular}[c]{@{}l@{}}anatips, bodygoals, bodyimage, bodypositivity, bonespo,\\ chloetingchallange, cleaneating, cleanvegan, dietculture, \\ eatingdisorder, edrecovery, edtwt, edvent, fatacceptance, \\ fatspo, fearfood, foodistheenemy, healthyliving, \\ intermittentfasting, iwillbeskinny, juicecleanse, keto, \\ ketodiet, losingweight, lowcalrestriction, m34nspo, \\ meanspo, midriff, ozempic, proana, proanatips, promia, \\ redbracetpro, semaglutide, skinnycheck, skinnydiet, \\ slimmingworld, sweetspo, thighgapworkout, \\ thinspo, thinspoa, watercleanse, wegovy, weightloss, \\ weightlossjourney, weightlossmotivation, weightlosstips, \\ whatieatinaday\end{tabular}} \\ \hline
\end{tabular}
\caption{List of keywords used to retrieve tweets. Terms are listed in alphabetical order.}
\label{tab:keywords}
\end{table}

\begin{figure*}[ht]
    \centering
\includegraphics[width=0.9\linewidth]{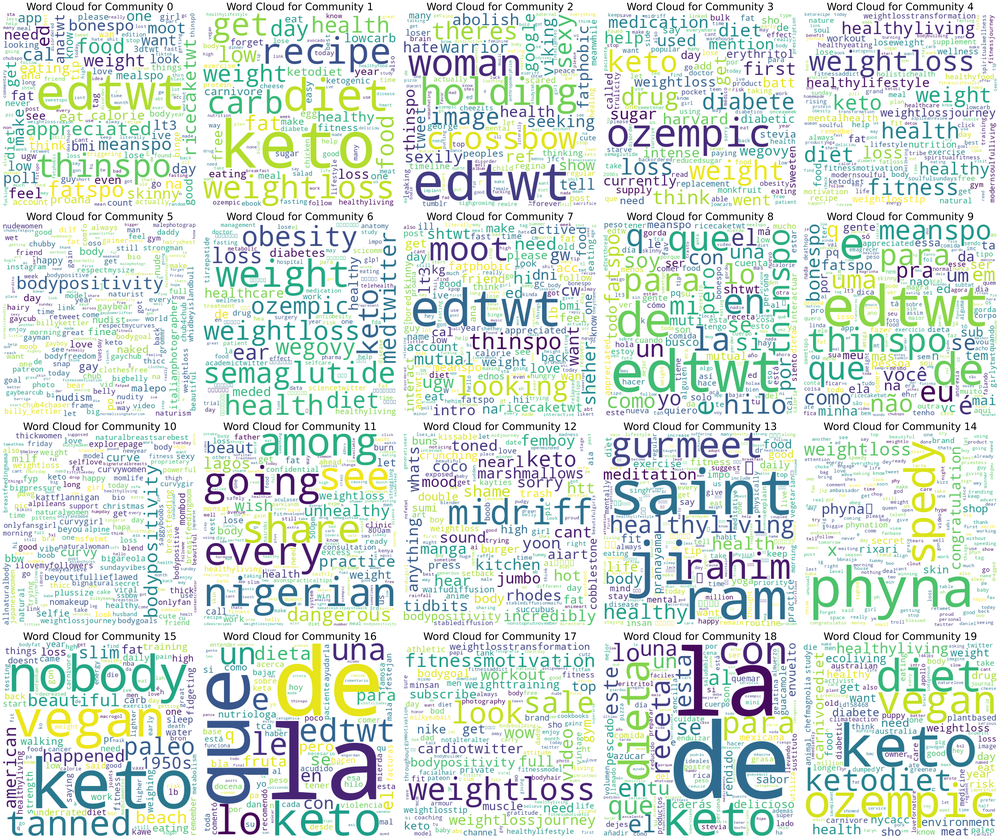}
    \caption{Wordclouds of popular terms appearing in the original tweets posted within each community.}
\label{fig:wordclouds}
\end{figure*}

\section{Identifying Online Eating Disorders Communities}

\subsection{Louvain Community Statistics}
The statistics of communities detected by Louvain community detection is shown in Table \ref{tab:rt_comm_stats}.

\begin{table*}[ht]
    \centering
\addtolength{\tabcolsep}{-3.3pt}
\begin{tabular}{lccccccccccc}
\hline
\textbf{Comm}      & \textbf{0} & \textbf{1} & \textbf{2} & \textbf{3} & \textbf{4} & \textbf{5} & \textbf{6} & \textbf{7} & \textbf{8} & \textbf{9} &  \\ \hline
\# of users & 61,954     & 24,400     & 21,887     & 20,631     & 9,901      & 9,031      & 9,000      & 8,084      & 7,702      & 7,020      &            \\
\# of tweets       & 805,249    & 112,674    & 32,883     & 37,788     & 193,348    & 24,395     & 21,369     & 82,702     & 70,764     & 71,970     &         \\ \hline
\textbf{Comm}      & \textbf{10} & \textbf{11} & \textbf{12} & \textbf{13} & \textbf{14} & \textbf{15} & \textbf{16} & \textbf{17} & \textbf{18} & \textbf{19} & \textbf{total} \\ \hline
\# of users &   6,477  & 6,158     &  5,181    &  4,528   &   3,682    &  3,672    &  3,360     &  3,163     &  3,086    &   2,865    &  221,887         \\
\# of tweets       &  15,796   & 9,254   & 7,019  &  103,177   &  260,971   &  5,338   & 4,881  &  5,065   &   4,612   &   7,021   & 1,876,276        \\ \hline
\end{tabular}
\addtolength{\tabcolsep}{3.3pt}
\caption{Number of users (community size) and tweets in the top 20 largest communities respectively and in total.}
\label{tab:rt_comm_stats}
\end{table*}

\section{LLM Finetuning}

\subsection{Demonstration Template for LLM finetuning}
The instructions for finetuning LLMs are shown in Table \ref{tab:inst}. For tweet generation demonstrations, each tweet is paired with a randomly sampled instruction from the table. An example prompt template is shown below.

\noindent \texttt{Instruction: You're part of the online community \{community\_name\} now. What would you tweet ?}\\
\noindent \texttt{Response: \{Tweet\}}

\subsection{Prompt Template for Tweet Generation by Finetuned LLM}
An example prompt template is shown below.

\noindent \texttt{Prompt: You're part of the online community \{community\_name\} now. What would you tweet about \{topic\}?}

\subsection{Prompt Template for Tweet Generation by Vanilla LLM}
An example prompt template is shown below.

\noindent \texttt{Prompt: You're part of the online community \{community\_name\}, which focuses on \{community description\} now. What would you tweet about \{topic\}?}

\subsection{Demonstration Template for Community Classification}
\noindent \texttt{Instruction: From these communities: Eating Disorder, Keto \& Diet, Body Image, Anti Eating Disorder, Healthy lifestyle \& Weight Loss, and Weight Loss Drugs, which community does this Tweet belong to? \{Tweet\}} \\ 
\noindent \texttt{Response: \{community\_name\}}

\begin{table*}[ht]
\centering
\begin{tabular}{cl}
\hline
\textbf{Index} & \multicolumn{1}{c}{\textbf{Instruction}} \\ \hline
\textbf{1} & \begin{tabular}[c]{@{}l@{}}You're part of the online community \{community\_name\} now. \\ What would you tweet ?\end{tabular} \\
\textbf{2} & \begin{tabular}[c]{@{}l@{}}If you were in the online community \{community\_name\},\\ what tweet would you send out?\end{tabular} \\
\textbf{3} & \begin{tabular}[c]{@{}l@{}}Say you're a member of the online community \{community\_name\}.\\ What's your tweet today?\end{tabular} \\
\textbf{4} & \begin{tabular}[c]{@{}l@{}}As a \{community\_name\} online community member,\\ what would you want to tweet about?\end{tabular} \\
\textbf{5} & \begin{tabular}[c]{@{}l@{}}You've joined the online community \{community\_name\}.\\ What's on your mind to tweet?\end{tabular} \\
\textbf{6} & \begin{tabular}[c]{@{}l@{}}Pretend you're hanging out in the online community \{community\_name\}.\\ What tweet would you drop?\end{tabular} \\
\textbf{7} & \begin{tabular}[c]{@{}l@{}}Imagine you're tweeting from the online community \{community\_name\}.\\ What would you say?\end{tabular} \\
\textbf{8} & \begin{tabular}[c]{@{}l@{}}You're now a voice in the online community \{community\_name\}.\\ What's your tweet?\end{tabular} \\
\textbf{9} & \begin{tabular}[c]{@{}l@{}}Step into the shoes of someone from the online community \{community\_name\}\\ and tweet something.\end{tabular} \\
\textbf{10} & \begin{tabular}[c]{@{}l@{}}If you were chatting as part of the online community \{community\_name\},\\ what would you tweet?\end{tabular} \\
\textbf{11} & \begin{tabular}[c]{@{}l@{}}You're now part of the \{community\_name\} online community.\\ What kind of tweet would you send out to engage with fellow members?\end{tabular} \\
\textbf{12} & \begin{tabular}[c]{@{}l@{}}Imagine you're an active participant in the \{community\_name\} online group.\\ Draft a tweet that captures the interests and spirit of the community.\end{tabular} \\
\textbf{13} & \begin{tabular}[c]{@{}l@{}}Put yourself in the shoes of someone really involved with the \{community\_name\} \\ online community. Craft a relatable tweet that resonates with members.\end{tabular} \\
\textbf{14} & \begin{tabular}[c]{@{}l@{}}You're deeply immersed in the \{community\_name\} online community discussions.\\ Share a tweet that sparks conversation on relevant topics.\end{tabular} \\
\textbf{15} & \begin{tabular}[c]{@{}l@{}}You're right in the mix of the digital sphere of the \{community\_name\} online group.\\ Compose a tweet that reflects the shared voice and passions.\end{tabular} \\
\textbf{16} & \begin{tabular}[c]{@{}l@{}}You're an influential voice within the \{community\_name\} online community.\\ Author an insightful tweet that inspires dialogue among members.\end{tabular} \\
\textbf{17} & \begin{tabular}[c]{@{}l@{}}You're respected as a thought leader engaging with the \{community\_name\} online community.\\ Tweet something that provokes intellectual discourse.\end{tabular} \\
\textbf{18} & \begin{tabular}[c]{@{}l@{}}You're entrenched in the activities of the \{community\_name\} online group.\\ Tweet an observation or perspective that contributes meaningfully.\end{tabular} \\
\textbf{19} & \begin{tabular}[c]{@{}l@{}}You're fully immersed in the virtual realm where \{community\_name\} members interact.\\ Craft a tweet that elevates the ongoing conversations.\end{tabular} \\
\textbf{20} & \begin{tabular}[c]{@{}l@{}}You're an esteemed voice that helps shape the \{community\_name\} online community.\\ Compose a tweet that encourages enriching engagement.\end{tabular} \\ \hline
\end{tabular}
\caption{Instructions used to finetune the LLMs with, where \emph{\{community name\}} is one of the six textual labels (from \emph{Pro Eating Disorder}, \emph{Healthy Lifestyle and Weight Loss}, \emph{Keto and Diet}, \emph{Weight Loss Drugs}, and \emph{Body Image}) assigned for the grouped communities as shown in table \ref{tab:comm_summaries}.
These prompts are designed to prompt the LLM to generate tweets.}
\label{tab:inst}
\end{table*}


\begin{table*}[h]
\centering
\begin{tabular}{|l|}
\hline
\multicolumn{1}{|c|}{\textbf{Keywords}}           \\ \hline
\textit{\begin{tabular}[c]{@{}l@{}} thinspo, fitspo, deathspo, caloric restriction,\\ caloric counting, purging, food rules, steroid, \\ meanspo, ozempic, wegovy,  fatspo, fatphobia, thighgap, \\ excessive exercising, body dysmorphia, working out\end{tabular}} \\ \hline
\end{tabular}
\caption{List of topics used to generate tweets.}
\label{tab:topics}
\end{table*}

\section{Screening Online Communities}
\subsection{Stanford-Washington University Eating Disorder (SWED) 3.0 Screener}

The 11 questions in the questionnaire are shown below.

\begin{enumerate}
    \item Are you currently in treatment for an eating disorder?
    \begin{enumerate}[label=(\alph*)]
        \item No
        \item Yes
        \item Not currently, but I have been in the past
    \end{enumerate}

    \item What was your lowest weight in the past year, including today, in pounds?

    \item What is your current weight in pounds?

    \item What is your current height in inches?

    \item How much more or less do you feel you worry about your weight and body shape than other people your age?
    \begin{enumerate}[label=(\alph*)]
        \item I worry a lot less than other people
        \item I worry a little less than other people
        \item I worry about the same as other people
        \item I worry a little more than other people
        \item I worry a lot more than other people
    \end{enumerate}

    \item How afraid are you of gaining 3 pounds?
    \begin{enumerate}[label=(\alph*)]
        \item Not afraid of gaining
        \item Slightly afraid of gaining
        \item Moderately afraid of gaining
        \item Very afraid of gaining
        \item Terrified of gaining
    \end{enumerate}

    \item When was the last time you went on a diet?
    \begin{enumerate}[label=(\alph*)]
        \item I have never been on a diet
        \item I was on a diet about one year ago
        \item I was on a diet about 6 months ago
        \item I was on a diet about 3 months ago
        \item I was on a diet about 1 month ago
        \item I was on a diet less than 1 month ago
        \item I’m on a diet now
    \end{enumerate}

    \item Compared to other things in your life, how important is your weight to you?
    \begin{enumerate}[label=(\alph*)]
        \item My weight is not important compared to other things in my life
        \item My weight is a little more important than some other things
        \item My weight is more important than most, but not all, things in my life
        \item My weight is the most important thing in my life
    \end{enumerate}

    \item Do you ever feel fat?
    \begin{enumerate}[label=(\alph*)]
        \item Never
        \item Rarely
        \item Sometimes
        \item Often
        \item Always
    \end{enumerate}

    \item In the past 3 months, how many times have you had a sense of loss of control AND you also ate what most people would regard as an unusually large amount of food at one time, defined as definitely more than most people would eat under similar circumstances?

    \item In the past 3 months, how many times have you done any of the following as a means to control your weight and shape:
    \begin{enumerate}[label=(\alph*)]
        \item Made yourself throw up?
        \item Used diuretics or laxatives?
        \item Exercised excessively? i.e. pushed yourself very hard; had to stick to a specific exercise schedule no matter what -- for example even when you were sick/injured or if it meant missing a class or other important obligation; felt compelled to exercise
        \item Fasted? i.e. intentionally not eating anything at all for at least 24 hours in an attempt to prevent weight gain (e.g., that is feared as a result of binge eating) or to lose weight
    \end{enumerate}

    \item Have you experienced significant weight loss (or are at a low weight for your age and height) but are not overly concerned with the size and shape of your body?
    \begin{enumerate}[label=(\alph*)]
        \item Yes
        \item No
    \end{enumerate}
\end{enumerate}

\subsection{Prompt Template for SWED Question Answering}
\noindent \texttt{Prompt: You’re now part of the Eating Disorder. \{question\}. Respond to the following question only with the letter at the beginning of each option or with a number.}



\end{document}